\documentclass[preprint,preprintnumbers,amsmath,amssymb,nofootinbib]{revtex4-2}
\usepackage{amsfonts}    
\usepackage{amssymb}
\usepackage{amsmath}
\usepackage{latexsym}
\usepackage{eepic}
\usepackage{graphicx}
\usepackage[usenames,dvipsnames]{color}
\usepackage{color}

\usepackage[active]{srcltx}
\usepackage{epstopdf}

\newcommand{\al}{\alpha}
\newcommand{\pa}{\partial}

\newcommand{\veps}{\varepsilon}

\newcommand{\la}{\lambda}

\newcommand{\om}{\omega}

\newcommand{\De}{\Delta}

\newcommand{\rar}{\rightarrow}

\newcommand{\non}{\nonumber}

\begin{document}

\title{Wolfes model {\it aka} $G_2/I_6$-rational integrable model: $g^{(2)}, g^{(3)}$ hidden algebras and quartic polynomial algebra of integrals}

\author{Juan Carlos Lopez Vieyra}
\email{vieyra@nucleares.unam.mx}
\author{Alexander~V.~Turbiner}
\email{turbiner@nucleares.unam.mx (Corresponding author)}
\affiliation{Instituto de Ciencias Nucleares, Universidad Nacional
Aut\'onoma de M\'exico, Apartado Postal 70-543, 04510 M\'exico,
D.F., Mexico{}}

\begin{abstract}
One-dimensional 3-body Wolfes model with 2- and 3-body interactions also known as $G_2/I_6$-rational integrable model of the Hamiltonian reduction is exactly-solvable and superintegrable. Its Hamiltonian $H$ and two integrals ${\cal I}_{1}, {\cal I}_{2}$, which can be written as algebraic differential operators in two variables (with polynomial coefficients) of the 2nd and 6th orders, respectively, are represented as non-linear combinations of $g^{(2)}$ or $g^{(3)}$ (hidden) algebra generators in a minimal manner. By using a specially designed MAPLE-18 code to deal with algebraic operators it is found that $(H, {\cal I}_1, {\cal I}_2, {\cal I}_{12} \equiv [{\cal I}_1, {\cal I}_2])$ are the four generating elements of the {\it quartic} polynomial algebra of integrals. This algebra is embedded into the universal enveloping algebra $g^{(3)}$. In turn, 3-body/$A_2$-rational Calogero model is characterized by cubic polynomial algebra of integrals, it is mentioned briefly.
\end{abstract}

\maketitle

\clearpage

\section{Introduction}

Remarkable 3-body Wolfes model with 2- and 3-body interactions \cite{Wolfes:1974}, see also \cite{Marchioro:1974}, also known as $G_2/I_6$-rational integrable model of the Hamiltonian reduction \cite{Ol-Per:1977}-\cite{OP:1983}, represents  three interacting bodies on the line.
If written in individual Cartesian coordinates $(x_1,x_2,x_3)$, it is described
by the Hamiltonian
\begin{equation}
\label{H}
{\cal H}\ \equiv\ H_{G_2/I_6}^{(\rm rational)}\ =\
-\frac{1}{2} \sum_{k=1}^{3}\bigg[\,{\frac {\pa^{2}}{\pa {{x}_k}^{2}}} + \om^2 x_k^2\bigg]
+ g_s\sum_{k<l}^{3}\frac{1}{(x_{k} - x_{l})^2}
+ 3 g_l\sum_{ k<l ,\ k,l \neq m}^{3} \frac{1}{(x_{k} + x_{l}-2 x_{m})^2}\ ,
\end{equation}
where $\om$ is a frequency and $g_s=\tilde\nu(\tilde\nu-1) > -
\frac{1}{4}$, $g_l=\tilde\mu (\tilde\mu -1) > - \frac{1}{4}$ are coupling constants
associated with the two-body and three-body interactions, which are defined
by {\it short} and {\it long} roots of the $g_2$-root space, where Weyl group $W_{G_2}$ acts.
If $g_l=0$ (or equivalently, $\tilde\mu=0,1$) the Wolfes model is reduced to the celebrated
3-body Calogero model (or, equivalently, $A_2$-rational integrable system of the Hamiltonian reduction, see for discussion \cite{OP:1983}).

If CM motion in (1) is separated out, the relative motion is two-dimensional and it is separated out in polar coordinates $(r, \varphi)$ \cite{Wolfes:1974}: the model becomes a particular case of the TTW system \cite{TTW:2009},
\begin{equation}
\label{Hk=3}
 H_k (r,\varphi;\om, \al, \beta)\ =\ -\pa_r^2 -
 \frac{1}{r}\pa_r - \frac{1}{r^2}\pa_{\varphi}^2  + \om^2 r^2 +
 \frac{\al k^2}{r^2 \cos^2 {k \varphi}} + \frac{\beta k^2}{r^2 \sin^2 {k \varphi}}\ ,
\end{equation}
for $k=3$, where $\al, \beta > - \frac{1}{4 k^2}$ are parameters related to $g_{s,l}$ in (\ref{H}).
The Hamiltonian (\ref{Hk=3}) is invariant with respect to the dihedral $I_6$ group transformations. The case $k=1$ corresponds to the superintegrable and exactly-solvable Smorodinsky-Winternitz model.
Let us note that non-trivial properties of the models (\ref{H}) and
(\ref{Hk=3}) are related with singular terms in the potential, thus, without a loss of generality we can put $\om=0$ in (\ref{H}) and (\ref{Hk=3}).

The Hamiltonian of the Wolfes model (1) can be written in the algebraic form as a differential operator with polynomial coefficients. In order to get this form a sequence of changes of variables should be realized. As the first step the variables $(x_1,x_2,x_3) \to (y_1,y_2,Y)$,
where the Center-of-Mass (CM) coordinate $Y=\sum_{k=1}^3x_k$ and the relative coordinates $y_i=x_i-\frac{1}{3}Y, i=1,2$ are introduced.
The $3D$ Laplacian $\De^{(3)} \equiv \sum_{i=1}^3\frac{\pa^{2}}{\pa x_i^{2}}$ in these coordinates takes the form,
\[
    \De^{(3)}\ =\ 3\ \pa_Y^2\ +\
    \frac{2}{3}\ \left(\frac{\pa^{2}}{\pa y_1^{2}}+\frac{\pa^{2}}{\pa y_2^{2}} -
    \frac{\pa^{2}}{\pa y_1 \pa y_2}\right)\ .
\]
Then, after separating out the CM coordinate $Y$, we make a change of variables to symmetric polynomials $(y_1,y_2) \to (x,y)$:
\begin{equation}
\label{xy-variables}
        x =-(y_1^2+ y_2^2 + y_1 y_2),\qquad y = - y_1 y_2 (y_1 + y_2)\ ,
\end{equation}
see \cite{RTC:1998}, coding the permutation invariance of the Hamiltonian into a new variables. Together with a gauge transformation,
\[
   h\ \equiv \ -2\ \De^{-\tilde\nu}\De_1^{-\tilde\mu}\,e^{-\frac{\om}{2} x}\ ({\cal H}-E_0)\ \De^{\tilde\nu}\De_1^{\tilde\mu}\,e^{\frac{\om}{2} x}\ ,
\]
where $\De=\prod_{i<j}^3|x_i-x_j|$ is Vandermonde determinant in $x$-Cartesian variables
and $\De_1=\prod_{i<j; \ i,j\neq k}|x_i+x_j-2x_k|$ is, in fact, Vandermonde determinant in $y$-variables, $y_1 + y_2 + y_3=0$, $E_0=\frac{3}{2}\,\om (1 + 2 \tilde \nu + 2\tilde \mu)$ is the ground state energy. The change of variables (\ref{xy-variables}) brings the Hamiltonian (1) already to an algebraic form of a differential operator with polynomial coefficients. However, further change of variables
\begin{equation}
\label{uv-coordinates}
  (x,y) \to (u=x,v=y^2)\ ,
\end{equation}
reduces the Hamiltonian (1) to its final algebraic form:
\begin{align}
  h_a \equiv h_{G_2/I_6}^{(r)}(u,v)\ &=\ u \frac{\pa^2}{\pa u^2}\ +\
    6 v \frac{\pa^2}{\pa u \pa v}\ -\ \frac{4 }{3} u^2 v \frac{\pa^2}{\pa v^2}\ +\
   (1+3\nu) \frac{\pa}{\pa u}\ -\ \frac{2}{3} u^2\frac{\pa}{\pa v} \non
\\[5pt] &
\label{hG2rat}
\hspace{30pt} +\ \la \left( 6\,{\frac {\pa}{\pa u}}  - 4\,{u}^{2}{\frac {\pa}{\pa v}}\right)
- 4\om u \frac {\pa}{\pa u} - 12\om v \frac {\pa}{\pa v}
\ ,
\end{align}
where the parameters $\la$ and $\nu$ are related to the coupling constants $g_{s,l}$ in (1) via either relations
\[
  \la = \frac{1}{3} \tilde\nu\ ,\ \nu=\tilde\mu + \frac{1}{3}\tilde\nu\ , \quad
  \mbox{or} \quad \la = \frac{1}{3} \tilde\mu\ ,\ \nu=\tilde\nu + \frac{1}{3}\tilde\mu\ ,
\]
which yields two dual ($\tilde\mu \longleftrightarrow \tilde\nu$) Hamiltonians $h_{G_2/I_6}^{(r,1)}(u,v) \longleftrightarrow h_{G_2/I_6}^{(r,2)}(u,v)\,$, see \cite{RTC:1998,Turbiner:2005}. Spectrum of the operator (\ref{hG2rat}) is given by
\[
    \veps_{n_1,n_2}\ =\ - 4\om (n_1 + 3 n_2)\ ,
\]
while the spectrum of the Hamiltonian (1) has the form \cite{RTC:1998}
\begin{equation}
\label{spectrum}
  E_{n_1,n_2}\ =\ -\frac{1}{2}\veps_{n_1,n_2} + E_0\ =\ 2\om (n_1 + 3 n_2) +
  \frac{3}{2}\,\om\, (1 + 2 \tilde \nu + 2\tilde \mu)\ ,
\end{equation}
where $n_{1,2}=0,1,2,\ldots $ are quantum numbers.

It can be immediately checked that the operator $h_{G_2/I_6}^{(r)}(u,v)$ (\ref{hG2rat}) has infinitely-many finite-dimensional invariant subspaces
\begin{equation}
\label{Pn}
   {\mathcal P}^{(s)}_n\ =\ <x^p y^q\ |\ 0 \leq p+sq \leq n >\ ,
\end{equation}
\[
   {h}_a:\,{\mathcal P}^{(s)}_n \ \rar {\mathcal P}^{(s)}_n \ ,\ n=0,1,2,\ldots \ ,
\]
for integer $s=2,3,4,\ldots $, for fixed $s$ they form the infinite flag,
\[
   {\mathcal P}^{(s)}_0 \subset {\mathcal P}^{(s)}_1 \subset \ldots \subset {\mathcal P}^{(s)}_n \subset \ldots  \equiv {\mathcal P}^{(s)}\ .
\]
It indicates to the very high degeneracy of the operator ${h}_a$.

It is easy to check that the ground-state eigenfunction in the space of relative motion
is given by
\begin{equation}
\label{psi0}
   \Psi_{0}(x)\ =\ (\De)^{\tilde \nu} (\De_1)^{\tilde \mu}
     e^{\frac{1}{2}\om  x} \ ,
\end{equation}
see \cite{Wolfes:1974}, which serves as a gauge factor in the gauge rotation
of the Hamiltonian (1). Furthermore, Wolfes showed explicitly \cite{Wolfes:1974}
that any eigenfunction of (1) can be written in the factorized form as
\begin{equation}
\label{psin}
     \Psi \ =\ \Psi_0(x)\, P_n(u,v) \ ,
\end{equation}
where $P_n(u,v) \in {\mathcal P}^{(s)}_n$ is a polynomial in two variables and $P_0(u,v)=1$. The first eigenpolynomials are written explicitly in \cite{RTC:1998}.

\section{Integrals}

It is well known \cite{OP:1983} that the $G_2/I_6$-rational model (1) is super-integrable, it has two integrals of motion, see for discussion \cite{TTW:2009}: one integral is of the second order (due to separation of variables in relative space polar coordinates) and another one is of the sixth order (inspired by the Hamiltonian reduction \cite{OP:1983}). It was shown in \cite{TTW:2009} that by making the gauge rotation with the ground state function (\ref{psi0}) as a gauge factor we will arrive at those integrals in the algebraic form if written in coordinates $(u, v)$ (\ref{uv-coordinates}), hence, to the form of differential operators with polynomial coefficients. As result of straightforward calculations the following algebraic operators occur:

(I) a  second-order integral  $x_{G_2}^{(r)} = x_3^{\rm TTW}(u,v)$, see \cite{TTW:2009}, after the gauge rotation it is given by
\begin{equation}
\label{x3uv}
 x_{G_2}^{(r)} (u,v)\ =\ \frac{4}{3}\,v \left( 4\,{u}^{3}+27\,v \right)
 {\frac {\pa ^{2}}{\pa {v}^{2}}}\ +\ \frac{4}{3} \left(  2 \left( 6\la +1  \right) \,{u}^{3}
 +   27 \,\left( 2\la + \nu\,+1 \, \right)\,v \right) {\frac {\pa}{\pa v}} \ ,
\end{equation}
which does not depend on $\om$. It can be checked that
\[
   x_{G_2}^{(r)}:\,{\mathcal P}^{(s)}_n \ \rar {\mathcal P}^{(s)}_n \ ,\ n=0,1,2,\ldots \ ,
\]
for integer $s=3,4,\ldots $. Thus, this operator preserves infinitely-many flags
${\mathcal P}^{(s)}, s=3,4,5,\ldots $. Note that to this integral it can be added arbitrary function of the algebraic Hamiltonian (\ref{hG2rat}) without loosing the integrability property.
In particular,
\begin{equation}
\label{I1-general}
  x_{G_2}^{(r)} \rar x_{G_2}^{(r)}\ +\ A\,  h_a\ ,
\end{equation}
where $A$ is parameter, it remains the second order differential operator.

(II) a sixth-order gauge rotated integral ${k}_{G_2}^{(r)}$ was written explicitly in \cite{TTW:2009} (see its ArXiv version~{\it v4}, eqs.(A1)-(A2)). In $(u,v)$ variables it has
the form
\begin{equation}
\label{k6uv}
      {k}_{G_2}^{(r)}(u,v)\ =\ ({{k}_{A_2}^{(r)}})^2 (u,v)\  +\ \sum_{p=1}^{5} \la^{p} {k}^{(p)}(u,v) \ ,
\end{equation}
where the first term is the square of the cubic integral of the $A_2$-rational integrable model $k_{A_2}^{(r)}(x,y)$, see for discussion \cite{ST:2015}, rewritten in $(u,v)$ coordinates.
The explicit expressions for $({{k}_{A_2}^{(r)}})^2 (u,v)$ and ${k}^{(p)}(u,v), p=1,2,3,4,5$
are presented in Appendix A. Note that this integral is defined ambiguously: it can be added arbitrary function of the algebraic Hamiltonian (\ref{hG2rat}) and the first integral (\ref{x3uv}) without loosing the integrability property.
In particular, if
\begin{equation}
\label{I2-general}
  k_{G_2}^{(r)} \rar k_{G_2}^{(r)}\ +\ B_1\,h_a^3 + B_2\,h_a^2\, x_{G_2}^{(r)} + B_3\,h_a\, \left(x_{G_2}^{(r)}\right)^2 + B_4\, \left(x_{G_2}^{(r)}\right)^3\ +
\end{equation}
\[
  +\ C_1\,h_a^2 + C_2\,h_a\, x_{G_2}^{(r)} + C_3\,\left(x_{G_2}^{(r)}\right)^2\ +\
  D_1\,h_a + D_2\,x_{G_2}^{(r)}\ ,
\]
where $\{B\}, \{C\}, \{D\}$ are parameters, the sixth order of the integral remains.
The operator $k_{G_2}^{(r)}$ preserves infinitely-many flags ${\mathcal P}^{(s)}, s=3,4,5,\ldots $.

Evidently, the Hamiltonian $h_{G_2/I_6}^{(r)}$ and the integral $x_{G_2}^{(r)}$, as well as
$h_{G_2/I_6}^{(r)}$ and $k_{G_2}^{(r)} $ have mutually common eigenfunctions. Furthermore, they preserve infinitely-many {\it common} flags ${\mathcal P}^{(s)}, s=3,4,5,\ldots $. Note that the minimal common flag for $h, x, k$ operators is ${\mathcal P}^{(3)}$.

\section{Polynomial algebra of integrals of $G_2/I_6$-rational model }

\subsection{The case of $G_2/I_6$-rational model}

For the sake of future convenience let us rename the Hamiltonian $h_{G_2/I_6}^{(r)}(u,v)$ (\ref{hG2rat}) and the integrals $x_{G_2}^{(r)}$ (\ref{x3uv}), $k_{G_2}^{(r)} $ (\ref{k6uv}) as
 \[
  h_{G_2/I_6}^{(r)}(u,v) \equiv H(u,v), \quad x_{G_2}^{(r)} (u,v) \equiv {\cal I}_1(u,v), \quad k_{G_2}^{(r)} (u,v) \equiv {\cal I}_2(u,v)\ .
 \]
By construction
\[
 [H,{\cal I}_1] = [H,{\cal I}_2] = 0\ .
\]
It is evident that the commutator of the integrals
\[
 [{\cal I}_1,{\cal I}_2] \equiv {\cal I}_{12}\ ,
\]
does not vanish, it remains the integral,
\begin{equation}
\label{I12}
  [H,{\cal I}_{12}]\ =\ 0\ .
\end{equation}
This is a rather complicated 7th order differential operator with polynomial coefficients. It is evident that
in spite of ambiguities in definition of ${\cal I}_1,{\cal I}_2$ (\ref{I1-general}), (\ref{I2-general}), the commutator  ${\cal I}_{12}$ is defined unambiguously.

After cumbersome and lengthy calculations by using specially designed MAPLE-18 code one can compute two commutators $[{\cal I}_1,{\cal I}_{12}]$ and  $[{\cal I}_2, {\cal I}_{12}]$ explicitly, which are the 8th order and the 12th order differential operators with $(u,v)$-dependent coefficient functions, respectively.
Surprisingly, they can be rewritten as very simple quartic polynomials in terms of
$H, {\cal I}_1, {\cal I}_2, {\cal I}_{12}$, see (\ref{hG2rat}), (\ref{x3uv}), (\ref{k6uv}) (!):
\[
 [{\cal I}_1,{\cal I}_{12}]\ =\ -\frac{32}{3} H^3 {\cal I}_1  - 144 {\cal I}_1 {\cal I}_2 - 72 {\cal I}_{12} -
 48 (2\la+\nu-1) \bigg(2 (6\la+1) H^3  + 27  (2\la+\nu+1) {\cal I}_2 \bigg)\ ,
\]
which is the 8th order differential operator, and
\[
 [I_2, {\cal I}_{12}]\ =\  \frac{32}{3}  H^3 {\cal I}_2 + 72 {\cal I}_2^2\ .
\]
which is the 12th order differential operator; the last commutator does not depend on ${\cal I}_1$ and ${\cal I}_{12}$, it appears in a certain minimal representation.

To summarize we have found that the operators $H, {\cal I}_1, {\cal I}_2, {\cal I}_{12}$ satisfy
the following algebraic relations:
\[
 [H,{\cal I}_{1}]  = 0\ ,\ [H,{\cal I}_{2}] = 0\ ,\ [I_{1},{\cal I}_{2}] = {\cal I}_{12}\ ,\ [H,{\cal I}_{12}]  = 0\ ,
\]
\[
 [{\cal I}_1,{\cal I}_{12}]  = -\frac{32}{3} H^3 {\cal I}_1  - 144 {\cal I}_1 {\cal I}_2 - 72 {\cal I}_{12} -
 48 (2\la+\nu-1) \bigg(2 (6\la+1) H^3  + 27  (2\la+\nu+1) {\cal I}_2 \bigg)\ ,
\]
\begin{equation}
\label{quartic-PA}
 [{\cal I}_2, {\cal I}_{12}]  =  \frac{32}{3}  H^3 I_2 + 72 {\cal I}_2^2\ .
\end{equation}
It is worth emphasizing that the dependence on the coupling constants (via the parameters $(\la, \nu)$) appears in the commutator $[{\cal I}_1,{\cal I}_{12}]$ {\it only}.
Hence, we arrive at {\it quartic} two-parametric polynomial algebra generated by the elements
$(H, {\cal I}_1, {\cal I}_2, {\cal I}_{12})$. At $\la=0$ this algebra is reduced to the
{\it quartic} one-parametric polynomial algebra of integrals which corresponds
to the $A_2$ rational Calogero model. In general, this algebra consists of all ordered monomials in the elements $H, {\cal I}_1, {\cal I}_2, {\cal I}_{12}$.

\subsection{The case of the $A_2$-rational model}

The Hamiltonian of the $A_2$-rational model corresponds to (1) with $g_l=0$. Separating the center-of-mass, then making the parametrization of the space of relative motion by $(x,y)$-coordinates (\ref{xy-variables}), after the gauge rotation with the ground state function (\ref{psi0}) at $\mu=0$ (as the gauge factor) we arrive at the algebraic operator, see e.g. \cite{TTW:2009} or \cite{ST:2015}. In $(x,y)$-coordinates this operator looks like,
\begin{equation}
\label{hA2rat}
   h_{A_2}^{(r)}(x,y)\ =\
   x \frac{\pa^2}{\pa x^2} + 3 y  \frac{\pa^2}{\pa x \pa y}  -  \frac{1}{3} x^2 \frac{\pa^2}{\pa y^2}
   +(1+3\nu) \frac{\pa}{\pa x} \ ,
\end{equation}
cf. (\ref{hG2rat}). In similar way one can get the gauge rotated 2nd order integral \cite{TTW:2009}
\begin{equation}
\label{xA2rat}
    x^{(r)}_{A_2}(x,y)\ =\
     \frac{1}{3}\left( 4\,{x}^{3} + 27\,{y}^{2} \right) {\frac {\pa^{2}}{\pa {y}^{2}}}
     + 9\,y \left(1 + 2\,\nu \right) {\frac {\pa }{\pa y}}\ ,
\end{equation}
cf. (\ref{x3uv}) and the gauge rotated 3rd order integral, see e.g. \cite{ST:2015},
\begin{align}
\label{kA2rat}
{k}_{A_2}^{(r)}(x,y)\ &=\
y{\frac {\partial ^{3}}{\partial {x}^{3}}}
- \frac{2}{3} \,{x}^{2}{\frac {\partial ^{3}}{\partial {x}^{2} \partial y}}
 - x y {\frac {\partial ^{3}}{\partial x \partial {y}^{2}}}
 - \left( {y}^{2} + {\frac {2}{27}\,{x}^{3}} \right) {\frac {\partial ^{3}}{\partial {y}^{3}}} \\ &
- \frac{2}{3}\,x \left( 2+3\,\nu \right) {\frac {\partial ^{2}}{\partial x\partial y}}
- y \left( 2+3\,\nu \right) {\frac {\partial ^{2}}{\partial {y}^{2}}}
- \frac{2}{9} \left( 2 + 3\,\nu \right)  \left( 1+3\,\nu \right) {\frac {\partial }{\partial y}} \,, \non
\end{align}
cf. (\ref{k6uv}).

Let us rename the Hamiltonian $h_{A_2}^{(r)}(x,y)$ and the integrals $x_{A_2}^{(r)} (x,y), k_{A_2}^{(r)}(x,y)$ as
 \[
  h_{A_2}^{(r)}(x,y) \equiv H(x,y), \quad x_{A_2}^{(r)}(x,y) \equiv {\cal I}_1(x,y), \quad k_{A_2}^{(r)}(x,y) \equiv {\cal I}_2(x,y)\ .
 \]
By construction
\[
 [H,{\cal I}_1] = [H,{\cal I}_2] = 0\ ,
\]
while the commutator
\[
 [{\cal I}_1,{\cal I}_2] \equiv {\cal I}_{12}\ ,
\]
is an 4th order differential operator with polynomial coefficients. It commutes with the Hamiltonian, $[H,{\cal I}_{12}]=0$.  This operator can {\it not} be rewritten in terms of the operators $H$ and $I_1$. Let us note that if the Hamiltonian is fixed the ambiguity in definition of the integrals is considerably reduced in comparison with $G_2/I_6$-rational integrable model:
the 2nd order integral $x^{(r)}_{A_2}$ {\it only} is defined ambiguously,
\[
   {\cal I}_1 \rar {\cal I}_1 + A\,H\ ,
\]
cf. (\ref{I1-general}), where $A$ is parameter.

After cumbersome and lengthy calculations by using a specially designed MAPLE-18 code we were able to find the commutators $[{\cal I}_1,{\cal I}_{12}], [{\cal I}_2, {\cal I}_{12}]$ explicitly, which are the 5th order and the 6th order differential operators with $(x,y)$-dependent polynomial coefficient functions, respectively.
Surprisingly, they can be rewritten as very simple quadratic/cubic polynomials in
$H, {\cal I}_1, {\cal I}_2, {\cal I}_{12}$\,, respectively:
\[
      [{\cal I}_1, {\cal I}_{12}]\ =\
        36\,{\cal I}_1 {\cal I}_2\ -\ 18 {\cal I}_{12}\ +\ 81 (1-4 \nu^2) {\cal I}_2\ ,
\]
\[
      [{\cal I}_{2}, {\cal I}_{12}]\ =\ (8/3) H^3\ +\ 18 {\cal I}_{2}^2\ ,
\]
where the last commutator is defined unambiguously.

Finally, we arrive at the conclusion that the operators $H, {\cal I}_1, {\cal I}_2, {\cal I}_{12}$ satisfy the following algebraic relations:
\[
 [H,{\cal I}_{1}]  = 0\ ,\ [H,{\cal I}_{2}] = 0\ ,\ [I_{1},{\cal I}_{2}] = {\cal I}_{12}\ ,\ [H,{\cal I}_{12}]  = 0\ ,
\]
\[
 [{\cal I}_1,{\cal I}_{12}]\ =\
        36\,{\cal I}_1 {\cal I}_2\ -\ 18 {\cal I}_{12}\ +\ 81 (1-4 \nu^2) {\cal I}_2\ ,
\]
\begin{equation}
\label{cubic-PA}
 [{\cal I}_2, {\cal I}_{12}]\ =\ (8/3) H^3\ +\ 18 {\cal I}_{2}^2\ .
\end{equation}

It is worth emphasizing that the dependence on the coupling constant via parameter $\nu$ appears in the commutator $[{\cal I}_1,{\cal I}_{12}]$ {\it only}.
Hence, we arrive at {\it quadratic-cubic} one-parametric polynomial algebra of integrals (\ref{cubic-PA}) generated by the elements $(H, {\cal I}_1, {\cal I}_2, {\cal I}_{12})$. This quadratic/cubic algebra of integrals (\ref{cubic-PA}) is alternative to the {\it quartic} one-parametric polynomial algebra of integrals of the $A_2$ rational Calogero model (\ref{quartic-PA}) at $\la=0$.

\section{Hidden algebra of $G_2$ rational integrable system.}

For fixed $n$ the invariant subspace ${\mathcal P}^{(s)}_n$ (\ref{Pn}) coincides with the finite-dimensional representation space of the algebra $g^{(s)}$: the infinite-dimensional, finitely-generated algebra of differential operators in two variables. This algebra was introduced in \cite{RTC:1998} (see for a discussion \cite{Turbiner:2005},\cite{Turbiner:2013t}) in relation to the $G_2$-rational and $G_2$-trigonometric integrable systems of the Hamiltonian reduction. It is spanned by the Euler-Cartan generator
\begin{equation}
\label{J0-g2}
{\tilde {\cal J}}_0(n)\ =\ r\pa_r \  +\ s u\pa_u \ - \ n\ ,
\end{equation}
and the generators
\[
 {\cal J}^1\  =\  \pa_r\ ,\
 {\cal J}^2_n\  =\ r \pa_r\ -\ \frac{n}{3} \ ,\
 {\cal J}^3_n\  =\ s u\pa_u\ -\ \frac{n}{3}\ ,
\]
\begin{equation}
\label{gl2r}
       {\cal J}^4_n\  =\ r^2 \pa_r \  +\ s r u \pa_u \ - \ n r\ =\ r {\tilde {\cal J}}_0(n)\ ,
\end{equation}
\[
 {\cal R}_{0}\  = \ \pa_u\ ,\ {\cal R}_{1}\  = \ r\pa_u\ ,\ldots \ ,\ {\cal R}_{s}\  = \ r^{s}\pa_u\ ,\
\]

\[
 {\cal T}^{(s)}_0\ =\ u\pa_{r}^s \ ,\ {\cal T}^{(s)}_1\ =\ u\pa_{r}^{s-1} {\tilde {\cal J}}_0{(n)}\ ,
 \ \ldots\ ,
\]
\begin{equation}
\label{gl2t}
 {\cal T}^{(s)}_s\ =\
  u{\tilde {\cal J}}_0{(n)}\ ({\tilde {\cal J}}_0{(n)} + 1)\ldots ({\tilde {\cal J}}_0{(n)} +s - 1) = \ u {\tilde {\cal J}}_0{(n)}\ {\tilde {\cal J}}_0{(n-1)}\ldots({\tilde {\cal J}}_0{(n)} - s + 1)\ ,
\end{equation}
where $n$ is a parameter, which has a meaning of the mark of representation. If $n$ is a non-negative integer, the space ${\mathcal P}_n$ is the common invariant subspace for the generators (\ref{J0-g2}), (\ref{gl2r}), (\ref{gl2t}), where the algebra acts irreducibly.
It is worth noting that for $s=1$ the algebra $g^{(1)}$ coincides with algebra $gl(3)$!

\noindent
{\it Remark.}
Note that the generators (\ref{J0-g2}), (\ref{gl2r}) span the subalgebra
$gl(2) \ltimes {\it R}^{s+1} \in g^{(s)}$, discovered by Sophus Lie around 1880 \cite{Lie:1880} as the algebra of vector fields,  see \cite{gko:1992} (Cases 24, 28 with $r=s$), which was extended much later to first order differential operators in \cite{gko:1994}.

By direct calculation one can show that the Hamiltonian $h_{G_2/I_6}^{(r)}$ (\ref{hG2rat}) and the integrals $x_{G_2/I_6}^{(r)}, k_{G_2/I_6}^{(r)}$ (\ref{x3uv}), (\ref{k6uv}) can be rewritten in terms of the generators of the algebra $g^{(3)}$, where the "positive-root", raising generator ${\cal J}^4_n$ is not present. Similarly for the case of $A_2$-rational integrable system: the Hamiltonian $h_{A_2}^{(r)}$ (\ref{hA2rat}) and both integrals $x_{A_2}^{(r)}$ (\ref{xA2rat}), $k_{A_2}^{(r)}$  (\ref{kA2rat}) can be rewritten in terms of generators of
the algebra $g^{(1)}=gl(3)$, see e.g. \cite{ST:2015}, where the positive-root generators are not present.

\section{Conclusions}

Many years ago it was shown the celebrated two-dimensional Smorodinsky-Winternitz (SW) model, which is the exactly-solvable and superintegrable, has a {\it quadratic} polynomial algebra of integrals \cite{Bonatsos:1994}, for discussion see \cite{Miller:2010} and is characterized by the $gl(3)$ hidden algebra \cite{TTW:2001,TTW:2009}.
As an immediate application of the polynomial algebra of integrals Bonatsos et al \cite{Bonatsos:1994} had demonstrated that the knowledge of the integrals allows to construct a deformed oscillator algebra; by using this fact the authors were able to calculate the spectra of SW model, and some other two-dimensional superintegrable systems.

In this article we study the quantum $G_2$-rational super-integrable system of 3 identical bodies on the line with pairwise and three-body interactions. It was shown that the Hamiltonian, two integrals and the commutator of integrals generate a {\it quartic} two-parametric polynomial algebra of integrals. This polynomial algebra is a subalgebra of the universal enveloping algebra of the hidden algebra $g^{(3)}$ of the Hamiltonian. Vanishing the three-body interaction in the $G_2$ Hamiltonian system reduces it to the $A_2$-rational superintegrable and exactly-solvable system of 3 identical bodies on the line with pairwise interactions. In this reduction the quartic polynomial algebra remains as well as the hidden algebra $g^{(3)}$. It must be emphasized that the possibility to perform the so cumbersome symbolic computations by using the MAPLE-18 code is related mostly with the fact that the Hamiltonian and integrals admit the algebraic form.

Alternatively, for the $A_2$-rational super-integrable system it can be built a cubic polynomial algebra of integrals with hidden algebra $gl(3)$ behind. Present study complements the recent article \cite{TE:2023}, where the two-body Coulomb problem in the Sturm representation had led to a new, two-dimensional, exactly-solvable, superintegrable quantum system in curved space of non-constant curvature with a $g^{(2)}$ hidden algebra and a cubic polynomial algebra of integrals. It is definitely interesting to check the existence of a polynomial algebra of integrals (of finite degrees) for the general super-integrable TTW system with rational index $k$. It will be addressed elsewhere.

\section*{Author’s contributions}

All authors contributed equally to this work

\section*{Acknowledgments}

\noindent
This work is partially supported by CONACyT grant A1-S-17364 and DGAPA grant IN113022 (Mexico).
The authors thank A.M. Escobar Ruiz for interest to the work and helpful discussions. This work is dedicated to the memory of Professor Willard Miller, Jr.

\section*{DATA AVAILABILITY}

Data sharing is not applicable to this article as no new data were created or analyzed in this study.


\appendix

\section{Coefficient functions in (\ref{k6uv})}

The explicit form of the operators $({{k}_{A_2}^{(r)}})^2$ and  ${k}^{(p)}, p=1\ldots 5$ in (\ref{k6uv}) at $\om=0$ are

{\footnotesize
\begin{align*}
  ({k_{A_2}^{(r)}})^2(u,v)  & =
 v{\frac {\pa^{6}}{\pa {u}^{6}}}
-\frac{8}{3}\,{u}^{2}v{\frac {\pa ^{6}}{\pa {u}^{5}\pa v}}
+{\frac {8}{9}\,uv \left( 2\,{u}^{3}-9\,v \right) {\frac {\pa ^{6}}{\pa {u}^{4} \pa {v}^{2}}}  }
+{\frac {16}{27}\,{v}^{2} \left( 16\,{u}^{3}-27\,v \right) {\frac {\pa
^{6}}{\pa {u}^{3}\pa {v}^{3}}}  }
\\ & \hskip -1.cm
+\frac {16}{81}\,{u}^{2}{v}^{2} \left( 8\,{u}^{3}+189\,v \right) {\frac {\pa ^{6}}{\pa {u}^{2} \pa {v}^{4}}}
+\frac {64 }{27} \,u{v}^{3} \left( 2\,{u}^{3}+27\,v \right) {\frac {\pa^{6}}{\pa u \pa {v}^{5}}}
+\frac {64}{729}\,{v}^{3} \left( 4\,{u}^{6}+108\,{u}^{3}v+729\,{v}^{2}
 \right) {\frac {\pa ^{6}}{\pa {v}^{6}}}
 \\[10pt]  &
-\frac{2}{3}\,{u}^{2}{\frac {\pa^{5}}{\pa {u}^{5}}}
+{\frac {8}{9}\,u \left( {u}^{3} - 3\left(3\,\nu+8 \right) v \right) {\frac {\pa^{5}}{\pa {u}^{4} \pa v}}  }
+\frac {8}{9}\,v \left( 2\left(6\,\nu+17\right) {u}^{3}
- 9\left(3\,\nu + 8\right) v \right) {\frac {\pa^{5}}{\pa {u}^{3} \pa {v}^{2}}}
\\ &
+{\frac {16}{27}\,{u}^{2}v \left( 8\,{u}^{3}+ 9\left( 12\,\nu+37 \right) v
 \right) {\frac {\pa ^{5}}{\pa {u}^{2} \pa {v}^{3}}}}
+{\frac {32 }{81}\,u{v}^{2} \left(  \left( 12\,\nu+89 \right) {u}^{3}+
 27 \left( 15\,\nu + 52 \right) v \right) {\frac {\pa ^{5}}{
\pa u \pa {v}^{4}}} }
\\ &
+{\frac {32}{243}\,{v}^{2} \left( 20\,{u}^{6}+ 18 \left( 6\,\nu+43 \right) v{u}^{3}+ 243\left( 6\,\nu+25 \right) {v}^{2} \right)
{\frac {\pa ^{5}}{\pa {v}^{5}}}  }
\\[10pt] &
-\frac{2}{3}\,u \left( 3\,\nu +2  \right) {\frac {\pa ^{4}}{\pa {u}^{4}}}
 + \frac{4}{9} \left( 6 (3+2\nu) u^3 -  (9\nu (2\nu+11)+103)v\right)\,
  {\frac {\pa ^{4}}{\pa {u}^{3} \pa v}  }
 \\ &
+\frac {4}{27}\,{u}^{2} \left( 8\,{u}^{3}+ \left( 180\,{\nu}^{2}+1116\,\nu+
1345 \right) v \right) {\frac {\pa ^{4}}{\pa {u}^{2} \pa {v}^{2}}}
\\ &
+ \frac {16}{27}\,uv \left(  8\left( 3\,\nu+11 \right) {u}^{3} +
3 \left( 72\,{\nu}^{2} + 450\,\nu + 631 \right) v \right) {\frac {\pa ^{4}}{ \pa u \pa {v}^{3}}}
\\ &
+\frac {16}{243}\,v \left( 60\,{u}^{6}+ 2\left( 36\,{\nu}^{2}+711\,\nu+2168\right) v{u}^{3}
+ 27\left( 117\,{\nu}^{2} + 873\,\nu + 1529 \right) {v}^{2} \right)
{\frac{\pa^{4}}{\pa {v}^{4}}}
 \\[10pt] &
 -\frac{2}{9}\, \left( 3\,\nu +2  \right)  \left( 3\,\nu +1 \right)
{\frac{\pa^{3}}{\pa {u}^{3}}}
  +{\frac{8}{27}\,{u}^{2} \left( 45\,{\nu}^{2}+117\,\nu+73 \right)
{\frac{\pa^{3}}{\pa {u}^{2} \pa v}}  }
\\ &
  + {\frac {8}{27}\,u \left(  \left( 12\,\nu+29 \right) {u}^{3}+
  \left(108\,{\nu}^{3}+990\,{\nu}^{2}+2355\,\nu+1594 \right) v \right)
{\frac{\pa^{3}}{\pa u \pa {v}^{2}}}  }
 \\ &
 + \frac{8}{243} \left( 20 u^6 +  27 (108\nu^3+1080\nu^2+3210\nu+2957) v^2
 + 6(72\nu^2 + 612\nu + 1051) v u^3 \right)
{\frac{\pa^{3}}{\pa {v}^{3}}}
 \\[10pt] &
 +{\frac { 8 }{27}\left(3\,\nu + 2 \right)  \left( 3\,\nu+5 \right) \left( 6\,\nu+5 \right)u\,
{\frac {\pa^{2}}{\pa u\pa v}} }
\\ &
+ \frac{4}{81} \left((324\nu^4+4050\nu^3+14643\nu^2+20421\nu+9592) v +
(72\nu^2+317\nu+ 376) u^3 \right)
{\frac {\pa^{2}}{\pa {v}^{2}}}
\\ &
+ {\frac {4}{81}\, \left( 3\,\nu +2 \right) ^{2} \left( 3\,\nu +1 \right)
 \left( 6\,\nu + 11\right) {\frac {\pa }{\pa v}}  } 
\end{align*}
}


{\footnotesize
\begin{align*}
  k^{(1)}(u,v)  &=
 -4\,{u}^{2}
   {\frac {\pa^{5}}{\pa {u}^{5}}}
+ \frac{16}{3}\,u \left( {u}^{3}-3\,v \right)
\frac {\pa^{5}}{\pa {u}^{4}\pa v}
+16\,v \left( 2\,{u}^{3}- 3\,v \right)
\frac {\pa^{5}}{\pa {u}^{3}\pa {v}^{2}}
\\ &
+\frac {32}{27}\,{u}^{2} v \left( 8\,{u}^{3}+135\,v \right)
\frac {\pa^{5}}{\pa {u}^{2}\pa {v}^{3}}
+\frac {64}{27}\,u{v}^{2} \left( 13\,{u}^{3}+135\,v \right)
{\frac {\pa^{5}}{\pa u\pa {v}^{4}}}
\\ &
+\frac {128}{81} \,{v}^{2} \left( {u}^{3}+9\,v \right)  \left( 2\,{u}^{3}+27 \,v \right)
{\frac {\pa^{5}}{\pa {v}^{5}}}
%
%
\\[10pt]&
- 12\, \left( \nu + 1 \right) u
{\frac {\pa ^{4}}{\pa {u}^{4}}}
+  \frac{8}{9}\left(  4 \left( 9\,\nu + 14 \right) {u}^{3}- 9 \left( 7\, \nu  + 8 \right) v \right)
{\frac {\pa ^{4}}{\pa {u}^{3}\pa v}}
\\ &
+ \frac {64}{27}\,{u}^{2} \left( 4\,{u}^{3}+ 9 \left( 11\,\nu+20 \right) v \right)
{\frac {\pa ^{4}}{\pa {u}^{2}\pa {v}^{2}}}
\\ &
+  \frac {64}{27}\,u v \left(  4\left( 3\,\nu + 14 \right) {u}^{3}+ 27 \left( 11\,\nu+26 \right) v \right)
{\frac {\pa ^{4}}{\pa u\pa {v}^{3}}}
\\ &
+ \frac {32}{81}\,v \left( 32\,{u}^{6}+  6 \left( 41\,\nu+199 \right) v{u}^{3}
+ 81\left( 32\,\nu+97 \right) {v}^{2} \right)
{\frac {\pa ^{4}}{\pa {v}^{4}}}
\\[10pt] &
-\frac{4}{3} \left( 9\,{\nu}^{2} + 15\,\nu + 5 \right)
{\frac {\pa ^{3}}{\pa {u}^{3}}}
+ \frac{16}{9} \left( 45\,{\nu}^{2} +  129\nu + 85 \right)  {u}^{2}
{\frac {\pa ^{3}}{\pa {u}^{2}\pa v}}
\\ &
+ \frac {16}{27}\,u \left(  \left( 48\,\nu+119 \right) {u}^{3}+
3 \left( 270\, {\nu}^{2}+ 966 \,\nu+845 \right) v \right)
{\frac {\pa ^{3}}{\pa u\pa {v}^{2}}}
\\ &
+ \frac{32}{243} \, \left( 46\,{u}^{6} +  3\left( 72\,{\nu}^{2} + 864\,\nu + 1717 \right) v {u}^{3}
+ 243\, \left( 30\,{\nu}^{2} + 147\,\nu + 179 \right) {v}^{2} \right)
 {\frac {\pa ^{3}}{\pa {v}^{3}}}
\\[10pt] &
+ \frac{16}{3} \left( 18\,{\nu}^{3} + 66\,{\nu}^{2} + 72\,\nu + 25 \right) u
{\frac {\pa ^{2}}{\pa u\pa v}}
\\ &
+  \frac{8}{81} \left( 2 \left(48\,\nu+163\right)  \left(3\,\nu+5\right) {u}^{3}
+  9\left( 414\,{\nu}^{3}+2268\,{\nu}^{2} + 3962\,\nu + 2293 \right) v \right)
 {\frac {\pa ^{2}}{\pa {v}^{2}}}
\\[10pt] &
+ \frac{8}{27}(3\nu+2) (54 \nu^3+198\nu^2+189\nu+52)
{\frac {\pa }{\pa v}}\ ,
\end{align*}
}

{\footnotesize
\begin{align*}
k^{(2)}(u,v)  &=
 -24\,u{\frac {\pa ^{4}}{\pa {u}^{4}}}
 + \frac{16}{3} \left(2 u^3 -15 v \right) {\frac {\pa^{4}}{\pa {u}^{3}\pa v}}
 +\frac{16}{9} u^2 \left(8 u^3+69 v \right) {\frac {\pa^{4}}{\pa {u}^{2}\pa {v}^{2}}}
\\ &
 \mbox{}+ \frac{64}{9} uv \left(8 u^3 + 45 v^2\right) {\frac {\pa ^{4}}{\pa u\pa {v}^{3}}}
 + \frac{128}{27}  \left( 2 u^6 v + 37 u^3 v^2 + 135 v^3 \right)
 {\frac {\pa ^{4}}{\pa {v}^{4}}}
\\[10pt] &
 \mbox{} - 16 \left( 3\,\nu + 2 \right) {\frac {\pa ^{3}}{\pa {u}^{3}}}
\mbox{}+   \frac{32}{3} \left( 12\nu+11 \right) u^2 \,
{\frac {\pa ^{3}}{\pa {u}^{2}\pa v}}
\\ &
+ \frac{32}{3} \left( \left(4\nu+11\right) u^4  +  \left(63\nu+86\right) v u \right)\,
{\frac {\pa ^{3}}{\pa u\pa {v}^{2}}}
\\ &
\mbox{}+ \frac{32}{9}  \left( 4 u^6 + 2\left(28\nu+85\right) v u^3 + 27\left(16\nu+31\right) v^2\right)
{\frac {\pa ^{3}}{\pa {v}^{3}}}
\\[10pt] &
+  \frac{32}{3} \left(27\nu^2+51\nu+23\right) u \,{\frac {\pa ^{2}}{\pa u\pa v}}
\\ &
\mbox{} + \frac{16}{27} \left(  2 \left(36\nu^2+231\nu+275 \right) u^3
+  3\left(621\nu^2+1809\nu+1327 \right) v\right)\,
{\frac {\pa ^{2}}{\pa {v}^{2}}}
\\[5pt] &
+ \frac{16}{9} \left(135\nu^3+351\nu^2+273\nu+71\right) \,
{\frac {\pa }{\pa v}}\,,
\end{align*}
}

{\footnotesize
\begin{align*}
 k^{(3)}(u,v)  &=
  -48\,{\frac {\pa ^{3}}{\pa {u}^{3}}}
  -64\,{u}^{2}{\frac {\pa^{3}}{\pa {u}^{2}\pa v}}
  +   \frac{64}{3} u \left(u^3 - 15 v \right)\,{\frac {\pa^{3}}{\pa u\pa {v}^{2}}}
  + \frac{128}{27} u^3\left(2 u^3 + 33 v \right)\,{\frac {\pa^{3}}{\pa {v}^{3}}}
  \\ &
\mbox{}-192\,u{\frac {\pa ^{2}}{\pa u\pa v}}
+ \frac{32}{9}  \left(10\,\left( 3\nu+5\right) u^3 + 9 \left(16\nu - 1\right) v \right)
{\frac {\pa ^{2}}{\pa {v}^{2}}}
\mbox{}+ \frac{32}{3} \left(9\nu+5\right) \left(3\nu+1 \right)
{\frac {\pa }{\pa v}}\,,
\end{align*}
}

{\footnotesize
\begin{align*}
k^{(4)}(u,v)  &=
     -384\,u{\frac {\pa ^{2}}{\pa u\pa v}}
     +  \frac{64}{3} \left(2 u^3 - 33 v \right) {\frac {\pa ^{2}}{\pa {v}^{2}}}
        - 64 \left(3\,\nu + 7 \right) {\frac {\pa }{\pa v}} \,,
\end{align*}
}

{\footnotesize
\begin{align*}
k^{(5)}(u,v)  &=
 -384\,{\frac {\pa }{\pa v}}\,.
\end{align*}
}

\end{document}